\documentclass[12pt]{article}
\usepackage{amsmath,amssymb}
\usepackage{jmr}
\newcommand {\nuc}[2]{\mbox{${}^{#1}\rm #2$}}
\newcommand {\half}{\mbox{$\frac{1}{2}$}}
\newcommand {\smhalf}{\raisebox{0.4ex}{\mbox{$\scriptstyle\frac{1}{2}$}}}
\newcommand{\bra}[1]{\mbox{$\langle{#1}|$}}
\newcommand{\ket}[1]{\mbox{$|{#1}\rangle$}}
\begin{document}
\title{Quantum Logic Gates and\\ Nuclear Magnetic Resonance\\ Pulse Sequences}
\author{}
\maketitle

\begin{quotation}
\raggedright\noindent

J.~A. Jones\\
OCMS, New Chemistry Laboratory, South Parks Road, Oxford, OX1~3QT, UK,
and Centre for Quantum Computation, Clarendon Laboratory, Parks Road,
Oxford OX1~3PU, UK\\[0.2in]
R.~H. Hansen\\
Centre for Quantum Computation, Clarendon Laboratory, Parks Road,
Oxford OX1~3PU, UK\\[0.2in]
M.~Mosca,
Centre for Quantum Computation, Clarendon Laboratory, Parks Road,
Oxford OX1~3PU, UK
and Mathematical Institute, 24--29 St Giles', Oxford, OX1~3LB, UK\\[0.2in]

Correspondence should be addressed to J.~A. Jones at the New Chemistry
Laboratory, telephone +44 1865 275923, FAX +44 1865 275921, e-mail
jones@bioch.ox.ac.uk.\\[0.2in]

Running title: Quantum Logic Gates

\end{quotation}

\begin{abstract}
There has recently been considerable interest in the use of Nuclear
Magnetic Resonance (NMR) as a technology for the implementation of small
quantum computers.  These computers operate by the laws of quantum
mechanics, rather than classical mechanics, and can be used to implement
new quantum algorithms.  Here we demonstrate how NMR can in principle be
used to implement all the elements required to build quantum computers,
and briefly discuss the potential applications of insights from quantum
logic to the development of novel pulse sequences with applications in
more conventional NMR experiments.
\end{abstract}
\begin{quotation}
{}\noindent
\end{quotation}
\newpage

\section{Introduction}
It is well known that it is difficult to simulate the behaviour of a
quantum mechanical system with a classical computer.  The difficulty
arises because quantum systems are not confined to their eigenstates but
can in general exist in any superposition of them; thus the vector space
needed to describe the system is extremely large.  For example a spin
system comprising $N$ spin-\smhalf\ nuclei occupies a Hilbert space with
$2^N$ dimensions.  For this reason it is impractical to simulate the
behaviour of spin systems with more than about a dozen nuclei.

In 1982 Feynman~\cite{FEY82} reversed this observation, suggesting that
quantum mechanical systems have a potentially very large information
processing ability.  Thus it should be possible to build quantum
mechanical computers which utilise this ability to achieve a computing
power well beyond that of corresponding classical systems.  The theory
of such quantum computers is now fairly well understood, but it has
proved extremely difficult to actually build one.  Recently, however,
attempts to build computers based on the NMR properties of small
molecules have exhibited considerable success~\cite{COR96, COR97, COR97b,
GER97, KNI97, JON98a, CHU98a, CHU98b, JON98b, JON98c}.

In this paper we show how NMR can be used to implement all the
``components'' required to construct quantum computers, and draw
comparisons between quantum logic gates and more conventional NMR pulse
sequences.  The potential role of quantum computing as a source of new
insights into NMR is also briefly addressed.

\subsection{Bits and qubits}
The basic unit of information in a classical computer is the bit, which
can take one of two values, 0 and 1.  Bits are then connected together
by logic gates to form logic circuits, which can implement more complex
logic operations, such as addition.  Developments in classical computers
have been driven by developments in the design and construction of logic
gates, which have steadily become smaller, faster, and cheaper.  However
this process is beginning to reach a fundamental limit, as logic gates
are reduced in size to atomic dimensions.  Progress beyond this limit
will require a different approach.

One obvious possibility is to implement bits and logic gates using
atomic components.  A bit can be implemented using any two state device,
such as the two quantum states of a two level system.  For example the
two Zeeman levels, \ket{\alpha} and \ket{\beta}, of a \nuc{1}{H} nucleus
in a magnetic field can be naturally described as a bit.  Similarly a
spin-system containing $N$ distinct \nuc{1}{H} nuclei can be modelled as
a set of $N$ bits.  Traditionally the lower and upper energy levels are
referred to as \ket{0} and \ket{1} respectively.

The time-evolution of a spin-system system under some Hamiltonian is
described by a series of unitary transformations, and so is of necessity
reversible.  Hence any quantum mechanical computer can only implement
reversible operations, and must be built from reversible logic gates.
This is not an important restriction as it has been shown that
reversible logic gates can be used to simulate traditional gates, and
thus reversible computers are just as powerful as their irreversible
counterparts\cite{BEN73,FEY96}.

There is, however, much more to quantum computers than the
implementation of classical algorithms using reversible logic: quantum
computers are also capable of implementing new types of quantum
mechanical algorithms\cite{DEU92,EKE96,GRO97,CLE98}, with potentially
enormous powers.  This occurs because a two-level quantum system is not
confined to its two eigenstates, but can exist in superpositions of
these two states; that is the system is not confined to \ket{0} and
\ket{1}, but can exist in states such as
\begin{equation}
c_0\ket{0}+c_1\ket{1}
\label{eq:super}
\end{equation}
where $c_0$ and $c_1$ are complex numbers and $c_0^*c_0+c_1^*c_1=1$.  A
nucleus in such a state is not really in state 0 or state 1, but is in
both states simultaneously.  For this reason a two level quantum system
is more than a simple bit, and is better described as a quantum
mechanical bit, or qubit.  A spin-system with $N$ nuclei contains $N$
qubits, and can be in a superposition of up to $2^N$ states.  This
ability to be in a large number of states simultaneously gives quantum
computers an intrinsic parallelism, which is exploited in quantum
algorithms.

\subsection{Qubits and NMR spin states}
Traditional designs for quantum computers comprise $N$ two-level systems
which are coupled to one another and have some specific interaction with
the outside world (so that they can be monitored and controlled) but are
otherwise isolated.  NMR systems are, by contrast, rather different.  In
particular a typical NMR sample comprises not just one spin-system, but
a very large number of copies, one from each molecule in the sample.
Thus while quantum computers are usually described using Dirac's bra and
ket notation, NMR systems are described using a density matrix, usually
written in the product operator basis\cite{SOR83}.  While it is possible
to draw close analogies between the states of traditional quantum
computers and those of NMR systems, it is necessary to proceed with some
caution.

\subsubsection{One qubit states}
A single qubit can be in either of its two eigenstates, \ket{0} and \ket{1},
or in some linear superposition of them.  Such a state is most conveniently
written as a column vector in Hilbert space: for example the state described
in equation~\ref{eq:super} is written as
\begin{equation}
\ket{\psi}=\begin{pmatrix} c_0 \\ c_1 \end{pmatrix}.
\end{equation}
The corresponding density matrix
\begin{equation}
\rho=\ket{\psi}\bra{\psi}=\begin{pmatrix}
c_0^*c_0 & c_1^*c_0 \\
c_0^*c_1 & c_1^*c_1
\end{pmatrix}
\end{equation}
can be decomposed as a sum of the four Pauli basis states,
$\half E$, $I_x$, $I_y$, and $I_z$.

Consider first the eigenstates, \ket{0} and \ket{1}.  These correspond
to the density matrices
\begin{equation}
\ket{0}\bra{0}=\begin{pmatrix} 1 & 0 \\ 0 & 0 \end{pmatrix}=\half E+I_z
\label{eq:rho0}
\end{equation}
and $\ket{1}\bra{1}=\half E-I_z$ respectively.  Multiples of the unit matrix
can be added to density matrices at will without effecting any NMR observable
in any way, and so as far as any NMR experiment is concerned the density
matrix $I_z$ is equivalent to \ket{0}, while $-I_z$ is equivalent to \ket{1}.
This simple approach is not, however, applicable to larger spin systems.

Next consider superpositions, such as $(\ket{0}+\ket{1})/\sqrt{2}$,
with its corresponding density matrix
\begin{equation}
\begin{pmatrix} \smhalf & \smhalf \\ \smhalf & \smhalf \end{pmatrix}
=\half E+I_x.
\label{eq:rhox}
\end{equation}
Once again multiples of the unit matrix can be ignored, and so $I_x$ is
equivalent to $(\ket{0}+\ket{1})/\sqrt{2}$.  Similarly $\ket{0}+i\ket{1}$
is equivalent to $I_y$, while $\ket{0}-\ket{1}$ is equivalent to $-I_x$.
Just as qubit eigenstates are closely related to magnetizations,
superpositions are closely related to NMR coherences.

\subsubsection{Two qubit states}
While the relationship between qubit states and NMR states is simple for
one qubit (one spin systems), this relationship is much more complicated
in systems with two or more qubits.  Indeed the problem of creating NMR
states corresponding to multi-qubit eigenstates prevented progress in
the implementation of NMR quantum computers for many years.

Typically quantum algorithms start with all qubits in state \ket{0}, which
for a two-qubit computer is the state \ket{00}.  The corresponding
density matrix
\begin{equation}
\renewcommand{\arraystretch}{0.67}
\ket{00}\bra{00}=
\begin{pmatrix}
1 & 0 & 0 & 0 \\
0 & 0 & 0 & 0 \\
0 & 0 & 0 & 0 \\
0 & 0 & 0 & 0 
\end{pmatrix}
\label{eq:rho00}
\end{equation}
is quite different from the thermal equilibrium density matrix
\begin{equation}
\renewcommand{\arraystretch}{0.67}
I_z+S_z=
\begin{pmatrix}
1 & 0 & 0 & 0 \\
0 & 0 & 0 & 0 \\
0 & 0 & 0 & 0 \\
0 & 0 & 0 & -1 
\end{pmatrix}.
\end{equation}
Cory \emph{et al.} have shown how this problem can be overcome
\cite{COR96, COR97, COR97b}.  The ideal density matrix
(Eq.~\ref{eq:rho00}) can be decomposed as the sum of four product
operators:
\begin{equation}
\ket{00}\bra{00}=\half\left( \half E + I_z + S_z + 2I_zS_z \right),
\end{equation}
and this sum (ignoring multiples of the unit matrix as usual) can be
assembled using conventional NMR techniques.  An alternative approach,
due to Gershenfeld and Chuang\cite{GER97}, works by selecting four
states from within a set of spin states arising from a larger
spin-system.  With a careful choice of states it is possible to find
four levels whose relative populations correspond to those of
$\ket{00}\bra{00}$, and these levels can be used as a pseudo two-spin
system.  While this approach is elegant, it is difficult to apply in
practice and has not been widely used.  A third approach, called
temporal averaging\cite{KNI97}, is conceptually related to Cory's
approach, but uses phase cycling instead of field gradients to select
the desired state.

Superpositions can be treated in much the same way, but they are not directly
related to coherences in any very simple way.  For example consider the state
$(\ket{00}+\ket{01})/\sqrt{2}$, in which the first spin is in state \ket{0},
while the second spin is in a superposition of states.  The corresponding
density matrix can be directly decomposed:
\begin{equation}
\renewcommand{\arraystretch}{0.67}
\begin{pmatrix}
\smhalf & \smhalf & 0 & 0 \\
\smhalf & \smhalf & 0 & 0 \\
0 & 0 & 0 & 0 \\
0 & 0 & 0 & 0 
\end{pmatrix}
=\half\left( \half E + I_z + S_x + 2I_zS_x \right),
\end{equation}
but a more subtle approach is to note that $(\ket{00}+\ket{01})/\sqrt{2}$
can be written as a product of one-qubit states
\begin{equation}
\frac{\ket{00}+\ket{01}}{\sqrt{2}}=\frac{\ket{0}(\ket{0}+\ket{1})}{\sqrt{2}}.
\end{equation}
The corresponding density matrix can also be decomposed as a direct
product of equations \ref{eq:rho0} and \ref{eq:rhox}:
\begin{equation}
\begin{pmatrix} 1 & 0 \\ 0 & 0 \end{pmatrix}
\otimes
\begin{pmatrix} \smhalf & \smhalf \\ \smhalf & \smhalf \end{pmatrix}
=\left(\half E+I_z\right)\times\left(\half E+S_x\right)
=\half\left(\half E+I_z+S_x+2I_zS_x\right).
\end{equation}
Note that, unlike the one qubit case, a simple superposition does not
correspond directly to an NMR coherence, but rather to a complex mixture
of coherences and populations.  Fortunately it is rarely necessary to
directly consider issues of this kind, as such states can be easily obtained
from states like Eq.~\ref{eq:rho00}.

Finally we consider superpositions of the form $(\ket{00}+\ket{11})/\sqrt{2}$,
which cannot be broken down into a product of one qubit states (such states
are normally referred to as entangled states).  As such states cannot
be factored it is necessary to decompose the corresponding density
matrices directly.  In this case
\begin{equation}
\renewcommand{\arraystretch}{0.67}
\begin{pmatrix}
\smhalf & 0 & 0 & \smhalf \\
0 & 0 & 0 & 0 \\
0 & 0 & 0 & 0 \\
\smhalf & 0 & 0 & \smhalf
\end{pmatrix}
=\half\left(\half E + 2I_zS_z + 2I_xS_x - 2I_yS_y\right),
\end{equation}
which is a mixture of longitudinal two-spin order and $DQ_x$
double quantum coherence.

\subsection{Global phase shifts}
One important consequence of the density matrix description of NMR
quantum computers is that it is completely insensitive to global phase
shifts.  In general the wavefunction of any isolated system can be
multiplied by an arbitrary phase shift without any observable consequences,
that is the states \ket{\psi} and $\ket{\psi'}=e^{i\phi}\ket{\psi}$ are
indistinguishable.  Indeed the absolute value of $\phi$ is completely
meaningless, although it is possible to determine \emph{relative} values
of $\phi$ for otherwise indistinguishable states by interference
experiments.

In the density matrix description of a state such global phases are not
preserved, as
\begin{equation}
\ket{\psi'}\bra{\psi'}=e^{i\phi}\ket{\psi}\bra{\psi}e^{-i\phi}
=\ket{\psi}\bra{\psi}.
\end{equation}
Thus global phases have no discernable effect in any NMR experiment, and
can be completely ignored.  This is fortunate, as most NMR pulse sequences
create global phase shifts, as discussed below, but as such phase shifts
are truly \emph{global} they can be neglected.

\section{One qubit gates}
One qubit gates act to modify the spin state of a single nucleus, and thus
correspond to rotations in single-spin subspaces.  Any rotation of this
kind can be achieved using RF fields, and so these gates are relatively
straightforward.  The gates can be implemented using selective pulses, in
which case the gate is applied to a single nucleus, or using hard pulses,
in which case the gate is simultaneously applied to a large number of
separate nuclei.  In this latter case the gate is more properly considered
as a product of one qubit gates, one for each nucleus affected.

\subsection{The {\sc not} gate}
The simplest one qubit gate is the {\sc not} gate\cite{FEY96}, which
is well known
from classical computing (thus {\sc not} is a classical one bit gate, as
well as a one qubit gate).  This gate, which we shall call {\sf N},
implements the rotation
\begin{equation}
\begin{split}
\ket{0}&\stackrel{\sf N}{\longrightarrow}\ket{1}\\
\ket{1}&\longrightarrow\ket{0}
\end{split}
\end{equation}
This can be described more compactly using a transformation matrix
\begin{equation}
\raisebox{-7pt}{
\begin{picture}(40,20)
\put(0,10){\line(1,0){10}}
\put(10,0){\framebox(20,20){\sf N}}
\put(30,10){\line(1,0){10}}
\end{picture}
}
\quad = \quad
\begin{pmatrix} 0 & 1 \\ 1 & 0 \end{pmatrix}
\end{equation}
where the symbol on the left signifies a {\sc not} gate in a quantum
circuit\cite{DEU89}.  Clearly this gate can be implemented using
a $180^\circ I_x$ pulse, as
\begin{equation}
e^{-i\pi I_x}=\begin{pmatrix} 0 & -i \\ -i & 0 \end{pmatrix}
\end{equation}
and this matrix has the correct form up to a global phase change.  This
global phase term is irrelevant, as the overall phase is not an NMR
observable quantity, and thus a $180^\circ I_x$ pulse provides a good
implementation of a {\sc not} gate.

As the {\sc not} operation is simply an inversion, it is hardly surprising
that it is implemented by a $180^\circ$ inversion pulse.  It might seem
that any inversion pulse, such a $180^\circ I_y$, would be suitable, but
this is not the case.  For example
\begin{equation}
e^{-i\pi I_y}=\begin{pmatrix} 0 & -1 \\ 1 & 0 \end{pmatrix}
\end{equation}
performs the transformation
\begin{equation}
\begin{split}
\ket{0}&\longrightarrow\ket{1}\\
\ket{1}&\longrightarrow-\ket{0}
\end{split}
\end{equation}
This is not a simple inversion, as it also negates the sign of the \ket{1}
state.  This is not very important when the gate is applied to a system
in an eigenstate, but is important when the gate is applied to a
superposition.  Consider, for example, the superposition
$(\ket{0}+\ket{1})/\sqrt{2}$, for which
\begin{equation}
\ket{0}+\ket{1}\stackrel{\sf N}{\longrightarrow}\ket{0}+\ket{1},
\end{equation}
while a $180^\circ I_y$ pulse would give $-\ket{0}+\ket{1}$, 
a quite different state.  This should not be surprising, as superpositions
are closely related to NMR coherences, and $I_x$ and $I_y$ pulses can have
quite different effects, depending on the relative phases of the pulse and
the coherent state.

\subsection{The square-root of {\sc not}}
The square-root of {\sc not} gate, {\sf V}, is a purely quantum mechanical
gate, in that it has no classical equivalent.  As the name implies,
{\sf V} has the property
\begin{equation}
{\sf V}^2={\sf VV}={\sf N}
\end{equation}
and so an obvious implementation is a $90^\circ I_x$ pulse,
\begin{equation}
e^{-i\pi/2\, I_x}=\frac{1}{\sqrt{2}}
\begin{pmatrix} 1 & -i \\ -i & 1 \end{pmatrix}.
\end{equation}
Once again this is equal to the ``ideal'' form,
\begin{equation}
\frac{1}{2}\begin{pmatrix}1+i & 1-i \\ 1-i & 1+i \end{pmatrix},
\end{equation}
up to a global phase.

The effect of {\sf V} is to take an eigenstate to a superposition of
eigenstates.  For example
\begin{equation}
\ket{0}\stackrel{\sf V}{\longrightarrow}\left(\ket{0}-i\ket{1}\right)/\sqrt{2}.
\end{equation}
This emphasises the quantum mechanical nature of {\sf V}, as such
superpositions do not have classical equivalents.

Pulses with other flip angles can be treated in much the same way:
for example a $60^\circ I_x$ pulse is equivalent to a cube-root of
{\sc not} gate.  This is not particularly interesting for one qubit
gates, but becomes more interesting when comparing two qubit logic gates
with spin state selective excitation sequences.

\subsection{The Hadamard gate}
The square root of {\sc not} is not unique in converting eigenstates to
superpositions: any $90^\circ$ pulse will have a similar effect, as will 
a number of other pulse sequences.  One particularly interesting sequence
would be one corresponding to the Hadamard gate, {\sf H}, which performs
the rotation
\begin{equation}
\begin{split}
\ket{0}\stackrel{\sf H}{\longrightarrow}
\left(\ket{0}+\ket{1}\right)/\sqrt{2}\\
\ket{1}\longrightarrow\left(\ket{0}-\ket{1}\right)/\sqrt{2}
\end{split}
\end{equation}
This has two useful properties.  First, it takes \ket{0} to a completely
uniform superposition, that is to a state where the coefficients in front
of \ket{0} and \ket{1} are \emph{identical}.  Secondly, {\sf H} is self
inverse, so applying {\sf H} twice is equivalent to doing nothing.

{\sf H} can be implemented in NMR using an off-resonance pulse, as
\begin{equation}
e^{-i\pi(I_x+I_z)/\sqrt{2}}=\frac{1}{\sqrt{2}}
\begin{pmatrix} -i & -i \\ -i & i \end{pmatrix},
\end{equation}
which has the desired form, ignoring a global phase shift as usual.
Alternatively this can be implemented using a three pulse
sandwich\cite{ERN87}, such as
\begin{equation}
45\,I_y - 180\,I_x - 45\,I_{-y}.
\end{equation}

When implementing quantum algorithms on NMR quantum computers, it is
often easier to replace the Hadamard, {\sf H}, by the pseudo-Hadamard
operator, {\sf h}, which has the form
\begin{equation}
{\sf h}=\frac{1}{\sqrt{2}} \begin{pmatrix} 1 & 1 \\ -1 & 1 \end{pmatrix}.
\end{equation}
This operation takes \ket{0} to a uniform superposition of \ket{0} and
\ket{1}, just like {\sf H}, but it is \emph{not} self inverse.  This can
be implemented using a $90^\circ_y$ pulse.  In many algorithms, a pair
of {\sf H} gates can be replaced by one {\sf h} gate and one
${\sf h}^{-1}$ gate; this last gate is easily implemented as a
$90^\circ_{-y}$ pulse.
Considered as an NMR operator, the {\sf H} sequence performs the rotation
\begin{equation}
I_z\stackrel{\sf H}{\longrightarrow}I_x\stackrel{\sf H}{\longrightarrow}I_z
\end{equation}
while {\sf h} performs the more conventional rotation
\begin{equation}
I_z\stackrel{\sf h}{\longrightarrow}I_x\stackrel{\sf h}{\longrightarrow}-I_z
\end{equation}

\section{Two qubit gates}
All the gates described above are one qubit gates: in NMR terms they perform
rotations within the subspace corresponding to a single spin.  If implemented
using hard pulses they can in general effect several spins, but the overall
effect is that each spin rotates within its own subspace, and the operation
can be decomposed as the product of two or more one spin operations.  True
two qubit gates correspond to rotations within a subspace corresponding to
two spins, and cannot be decomposed into a set of one qubit gates.  These
gates lie at the heart of quantum computation, as they provide conditional
dynamics: that is, the state of one spin can become dependent on the state
of another spin.

\subsection{The controlled-{\sc not} gate}
The fundamental two qubit gate is the controlled-{\sc not} gate, which
applies a {\sc not} gate to one qubit (the ``target'') if another qubit
(the ``control'') is in state \ket{0}.  It can be described as follows:
\begin{equation}
\renewcommand{\arraystretch}{0.67}
\raisebox{-17pt}{
\begin{picture}(40,45)
\put(0,40){\line(1,0){40}}
\put(20,40){\line(0,-1){20}}
\put(20,40){\circle*{5}}
\put(0,10){\line(1,0){10}}
\put(10,0){\framebox(20,20){\sf N}}
\put(30,10){\line(1,0){10}}
\end{picture}}
=
\begin{pmatrix} 1 & 0 & 0 & 0 \\
                0 & 1 & 0 & 0 \\
                0 & 0 & 0 & 1 \\
                0 & 0 & 1 & 0 
\end{pmatrix}
\end{equation}
A variety of methods for implementing this gate have been described, but it
is more useful to consider a general approach to gates of this kind.  It is
well known in quantum computation that controlled gates, such as the
controlled {\sc not}, are related to controlled phase shifts by the
Hadamard transform.  For example a controlled-{\sc not} gate can be
replaced by the three gate network
\begin{equation}
\raisebox{-20pt}{
\begin{picture}(40,45)
\put(0,40){\line(1,0){40}}
\put(20,40){\line(0,-1){20}}
\put(20,40){\circle*{5}}
\put(0,10){\line(1,0){10}}
\put(10,0){\framebox(20,20){\sf N}}
\put(30,10){\line(1,0){10}}
\end{picture}}
\quad = \quad
\raisebox{-20pt}{
\begin{picture}(100,45)
\put(0,40){\line(1,0){100}}
\put(50,40){\line(0,-1){20}}
\put(50,40){\circle*{5}}
\put(0,10){\line(1,0){10}}
\put(10,0){\framebox(20,20){\sf H}}
\put(30,10){\line(1,0){10}}
\put(40,0){\framebox(20,20){\boldmath$\pi$}}
\put(60,10){\line(1,0){10}}
\put(70,0){\framebox(20,20){\sf H}}
\put(90,10){\line(1,0){10}}
\end{picture}}
\end{equation}
where {\sf H} are one qubit Hadamard gates, and \mbox{\boldmath$\pi$} is
given by
\begin{equation}
\renewcommand{\arraystretch}{0.67}
\mbox{\boldmath$\pi$}=\begin{pmatrix}
1 & 0 & 0 & 0 \\
0 & 1 & 0 & 0 \\
0 & 0 & 1 & 0 \\
0 & 0 & 0 & -1
\end{pmatrix}
\end{equation}
which performs the transformation $\ket{11}\rightarrow -\ket{11}$, while
leaving other basis states unchanged.

When implementing controlled gates in NMR pulse sequences, it is simpler to
replace the Hadamard gates by pseudo Hadamard gates, as described above.
Hence a general controlled gate can be implemented in NMR using the
network
\begin{equation}
\begin{picture}(100,45)(0,0)
\put(0,40){\line(1,0){100}}
\put(50,40){\line(0,-1){20}}
\put(50,40){\circle*{5}}
\put(0,10){\line(1,0){10}}
\put(10,0){\framebox(20,20){${\sf h}^{-1}$}}
\put(30,10){\line(1,0){10}}
\put(40,0){\framebox(20,20){\boldmath$\phi$}}
\put(60,10){\line(1,0){10}}
\put(70,0){\framebox(20,20){${\sf h}$}}
\put(90,10){\line(1,0){10}}
\end{picture}
\label{eq:hphih}
\end{equation}
where \mbox{\boldmath$\phi$} is a general controlled phase shift
\begin{equation}
\renewcommand{\arraystretch}{0.67}
\mbox{\boldmath$\phi$}=\begin{pmatrix}
1 & 0 & 0 & 0 \\
0 & 1 & 0 & 0 \\
0 & 0 & 1 & 0 \\
0 & 0 & 0 & e^{i\phi}
\end{pmatrix}
\label{eq:phi}
\end{equation}
which performs $\ket{11}\rightarrow e^{i\phi}\ket{11}$, while
leaving other basis states unchanged.

This general form for controlled gates in NMR should not be a surprise, as
there is a close link with composite $z$-pulses.  It is well known that
$z$-pulses can be replaced by three pulse sandwiches\cite{ERN87,FRE81};
for example a
$\theta_z$ pulse can be replaced by $90^\circ_{-x}\theta_{y}90^\circ_{x}$.
Similarly, by cyclic permutation of axes, a $\theta_x$ pulse can be
replaced by $90^\circ_{-y}\theta_{z}90^\circ_{y}$, which is equivalent to
${\sf h^{-1}}\theta_z{\sf h}$.  Since a {\sc not} gate (a $180^\circ_x$
pulse) can be implemented as an inverse pseudo Hadamard, followed by a
$180^\circ$ phase shift, followed by a pseudo Hadamard, it is hardly surprising
that a controlled-{\sc not gate} can be implemented in much the same way,
but using a controlled phase shift.

\subsection{Controlled phase shifts}
Controlled phase shifts, such as \mbox{\boldmath$\phi$} (Eq.~\ref{eq:phi}),
are relatively simple to implement, as they can always be decomposed
as a product of diagonal operators.  For example
\begin{equation}
\mbox{\boldmath$\phi$}=
\exp\left[-i\times\smhalf\phi\times\left(
-(\smhalf E)+I_z+S_z-2I_zS_z \right)\right].
\end{equation}
The last three terms are straightforward, but the first term is difficult
to obtain as it requires
a Hamiltonian proportional to $\smhalf E$.  This is not, however,
important, as this term simply imposes a global phase shift, and as such
can be ignored.  For the remaining three terms, $2I_zS_z$ is proportional
to the scalar coupling Hamiltonian, while $I_z$ and $S_z$ can be
implemented as periods of free precession or by using composite $z$-pulses.

The matrix \mbox{\boldmath$\pi$}, which lies at the heart of
the controlled-{\sc not} gate, can be implemented as 
\begin{equation}
(90^\circ\,I_z)(90^\circ\,S_z)(-90^\circ\,2I_zS_z)
\end{equation}
which can itself be achieved in a variety of ways.  The three terms above
commute, and so the three Hamiltonians can be applied in any order, while
the $I_z$ and $S_z$ terms can be implemented by free precession or by
any of a wide variety of different composite pulses.

Just like simple one qubit gates, two qubit controlled gates can also introduce
global phase shifts, but as long as these are global and universal, that
is they are applied to the the whole wavefunction and not just to the spins
participating in the gate and they are applied irrespective of the state
of the control bit, such phase shifts can be ignored.  This is indeed the
case: conceptually these phase shifts can be thought of as arising from
the lack of a $\smhalf E$ term in controlled phase shift gates, and thus
have the desired properties.

\subsection{The controlled square-root of {\sc not}}
Implementation of the controlled square-root of {\sc not} gate is simple
using the approach outlined in equations \ref{eq:hphih} and \ref{eq:phi},
with $\phi=\pi/2$.  The controlled phase shift is simply implemented as
\begin{equation}
(45^\circ\,I_z)(45^\circ\,S_z)(-45^\circ\,2I_zS_z).
\end{equation}
Note that there is a close relationship between this gate and the spin-state
selective excitation sequences\cite{MEI97a,MEI97b} which have been suggested
as a method for simplifying E-COSY spectra.  Clearly any other spin-state
selective pulse can be created in much the same way.

\section{Three qubit gates}
A wide variety of three bit gates have been investigated, but we will
confine our discussions to the Toffoli gate\cite{FEY96}, or
controlled-controlled-{\sc not}.  This takes the form
\begin{equation}
\raisebox{-27pt}{
\begin{picture}(40,65)
\put(0,60){\line(1,0){40}}
\put(20,60){\line(0,-1){20}}
\put(20,60){\circle*{5}}
\put(0,40){\line(1,0){40}}
\put(20,40){\line(0,-1){20}}
\put(20,40){\circle*{5}}
\put(0,10){\line(1,0){10}}
\put(10,0){\framebox(20,20){\sf N}}
\put(30,10){\line(1,0){10}}
\end{picture}}
\quad = \quad
\renewcommand{\arraystretch}{0.67}
\begin{pmatrix}
                1 & 0 & 0 & 0 & 0 & 0 & 0 & 0 \\
                0 & 1 & 0 & 0 & 0 & 0 & 0 & 0 \\
                0 & 0 & 1 & 0 & 0 & 0 & 0 & 0 \\
                0 & 0 & 0 & 1 & 0 & 0 & 0 & 0 \\
                0 & 0 & 0 & 0 & 1 & 0 & 0 & 0 \\
                0 & 0 & 0 & 0 & 0 & 1 & 0 & 0 \\
                0 & 0 & 0 & 0 & 0 & 0 & 0 & 1 \\
                0 & 0 & 0 & 0 & 0 & 0 & 1 & 0 
\end{pmatrix}
\label{eq:toffoli}
\end{equation}
and plays a central role in the theory of classical reversible computation
as it can be shown to be universal (that is, any reversible classical logic
circuit can be constructed entirely out of classical Toffoli gates).

It might seem that this gate could be implemented using double-controlled
phase shifts,
\begin{equation}
\raisebox{-27pt}{
\begin{picture}(40,65)(0,0)
\put(0,60){\line(1,0){40}}
\put(20,60){\line(0,-1){20}}
\put(20,60){\circle*{5}}
\put(0,40){\line(1,0){40}}
\put(20,40){\line(0,-1){20}}
\put(20,40){\circle*{5}}
\put(0,10){\line(1,0){10}}
\put(10,0){\framebox(20,20){\sf N}}
\put(30,10){\line(1,0){10}}
\end{picture}}
\quad = \quad
\raisebox{-27pt}{
\begin{picture}(100,65)(0,0)
\put(0,60){\line(1,0){100}}
\put(50,60){\line(0,-1){20}}
\put(50,60){\circle*{5}}
\put(0,40){\line(1,0){100}}
\put(50,40){\line(0,-1){20}}
\put(50,40){\circle*{5}}
\put(0,10){\line(1,0){10}}
\put(10,0){\framebox(20,20){${\sf h}^{-1}$}}
\put(30,10){\line(1,0){10}}
\put(40,0){\framebox(20,20){\boldmath$\pi$}}
\put(60,10){\line(1,0){10}}
\put(70,0){\framebox(20,20){${\sf h}$}}
\put(90,10){\line(1,0){10}}
\end{picture}}
\end{equation}
but while the above circuit is indeed correct this approach is not practical.
The double-controlled phase shift matrix, \mbox{\boldmath$\pi$}
can be decomposed as
\begin{equation}
\mbox{\boldmath$\pi$}=
\exp\left[-\frac{i\pi}{2}\left(
\smhalf E-I_z-R_z-S_z+2I_zR_z+2I_zS_z+2R_zS_z-4I_zR_zS_z
\right)\right]
\end{equation}
but this decomposition cannot be used as a guide to implementing
\mbox{\boldmath$\pi$}, as there is no NMR Hamiltonian directly corresponding
to $4I_zR_zS_z$.  For the same reason it is not possible to directly
implement a double-controlled square-root of {\sc not}, that is a doubly
spin state selective excitation sequence.

It is, however, possible to implement these gates by using more complex
networks of logic gates.  Indeed it has been shown that the combination
of a controlled-{\sc not} gate and a set of general one bit gates is
universal\cite{BAR95}, so that any other gate can be constructed from them.
This process is even simpler if the set of basic gates is slightly expanded;
for example a Toffoli gate can be implemented using the network
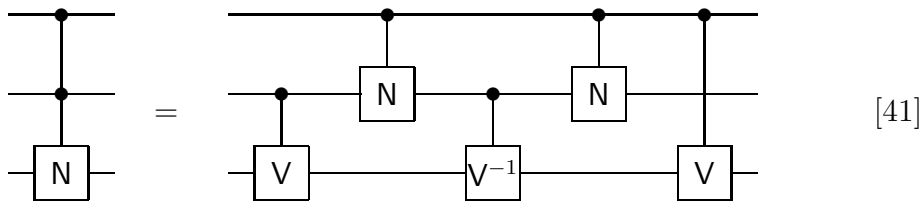
\begin{equation}
\raisebox{-30pt}{
\begin{picture}(40,75)
\put(0,70){\line(1,0){40}}
\put(20,70){\line(0,-1){30}}
\put(20,70){\circle*{5}}
\put(0,40){\line(1,0){40}}
\put(20,40){\line(0,-1){20}}
\put(20,40){\circle*{5}}
\put(0,10){\line(1,0){10}}
\put(10,0){\framebox(20,20){\sf N}}
\put(30,10){\line(1,0){10}}
\end{picture}}
\quad = \quad
\raisebox{-30pt}{
\begin{picture}(200,75)
\put(0,70){\line(1,0){40}}
\put(0,40){\line(1,0){40}}
\put(20,40){\line(0,-1){20}}
\put(20,40){\circle*{5}}
\put(0,10){\line(1,0){10}}
\put(10,0){\framebox(20,20){\sf V}}
\put(30,10){\line(1,0){10}}
\put(40,70){\line(1,0){40}}
\put(60,70){\line(0,-1){20}}
\put(60,70){\circle*{5}}
\put(40,40){\line(1,0){10}}
\put(50,30){\framebox(20,20){\sf N}}
\put(70,40){\line(1,0){10}}
\put(40,10){\line(1,0){40}}
\put(80,70){\line(1,0){40}}
\put(80,40){\line(1,0){40}}
\put(100,40){\line(0,-1){20}}
\put(100,40){\circle*{5}}
\put(80,10){\line(1,0){10}}
\put(90,0){\framebox(20,20){${\sf V}^{-1}$}}
\put(110,10){\line(1,0){10}}
\put(120,70){\line(1,0){40}}
\put(140,70){\line(0,-1){20}}
\put(140,70){\circle*{5}}
\put(120,40){\line(1,0){10}}
\put(130,30){\framebox(20,20){\sf N}}
\put(150,40){\line(1,0){10}}
\put(120,10){\line(1,0){40}}
\put(160,70){\line(1,0){40}}
\put(180,70){\line(0,-1){50}}
\put(180,70){\circle*{5}}
\put(160,40){\line(1,0){40}}
\put(160,10){\line(1,0){10}}
\put(170,0){\framebox(20,20){\sf V}}
\put(190,10){\line(1,0){10}}
\end{picture}}
\end{equation}

It should be noted that while it will be difficult to construct a true
Toffoli gate it is relatively simple to construct an approximate Toffoli 
gate, whose transformation matrix has the same underlying form as
equation~\ref{eq:toffoli}, but where the non-zero matrix elements are
not all equal to unity.  Such a gate was demonstrated early
on~\cite{COR96}, and can be used instead of a true gate in some
situations where it is the last gate in the logic network, such as error 
correction~\cite{COR98}.

\section{Applying logic networks in NMR}
Having shown that NMR pulse sequences can be used to implement all the basic
logic gates needed both for classical reversible computation and for quantum
computation, it is instructive to consider whether any more complex logic
networks might correspond to interesting NMR pulse sequences.  One obvious
candidate is the double-controlled square-root of {\sc not} circuit, which
corresponds to a doubly spin-state selective excitation sequence.  More
generally this family of gates involves the creation of an effective
Hamiltonian containing a term proportional to $4I_zR_zS_z$, which is not
easily accessible by conventional means, and may be useful in the generation
of multiple quantum coherences.

Another logic network which might prove useful is the {\sc swap} network
\begin{equation}
\raisebox{-20pt}{
\begin{picture}(50,45)
\put(0,10){\line(1,0){10}}
\put(0,40){\line(1,0){10}}
\put(10,10){\line(1,1){30}}
\put(10,40){\line(1,-1){30}}
\put(40,10){\line(1,0){10}}
\put(40,40){\line(1,0){10}}
\end{picture}
}
\quad = \quad
\raisebox{-20pt}{
\begin{picture}(120,50)
\put(0,40){\line(1,0){40}}
\put(20,40){\line(0,-1){20}}
\put(20,40){\circle*{5}}
\put(0,10){\line(1,0){10}}
\put(10,0){\framebox(20,20){\sf N}}
\put(30,10){\line(1,0){10}}
\put(40,40){\line(1,0){10}}
\put(50,30){\framebox(20,20){\sf N}}
\put(70,40){\line(1,0){10}}
\put(40,10){\line(1,0){40}}
\put(60,10){\line(0,1){20}}
\put(60,10){\circle*{5}}
\put(80,40){\line(1,0){40}}
\put(100,40){\line(0,-1){20}}
\put(100,40){\circle*{5}}
\put(80,10){\line(1,0){10}}
\put(90,0){\framebox(20,20){\sf N}}
\put(110,10){\line(1,0){10}}
\end{picture}
}
\quad = \quad
\renewcommand{\arraystretch}{0.67}
\begin{pmatrix}
1 & 0 & 0 & 0 \\
0 & 0 & 1 & 0 \\
0 & 1 & 0 & 0 \\
0 & 0 & 0 & 1
\end{pmatrix}
\end{equation}
which completely interchanges the states of the two spins involved.  Clearly
this is closely related to sequences like the double INEPT transfer step
used in sensitivity enhanced HSQC experiments\cite{CAV90,PAL91,KAY92}.
Unlike conventional
heteronuclear transfer steps, however, this sequence preserves the
states of both spins, by performing a complete swap.

More speculatively it may be possible to use quantum error correction
codes\cite{SHO95,STE96,STE97}
to reduce the effects of spin--spin relaxation upon NMR spectra.
Initial experiments in this direction\cite{COR98}
suggest that error correction does
work in NMR experiments, but that it is unlikely to have much practical
significance; this assessment may, however, prove too pessimistic.

\section{Acknowledgments}
This is a contribution from the Oxford Centre for Molecular Sciences which
is supported by the UK EPSRC, BBSRC and MRC.  JAJ thanks C.~M. Dobson
(OCMS) for his encouragement.  RHH thanks the Danish Research Academy
for financial assistance.  MM thanks CESG (UK) for their support.

\end{document}